\documentclass[twoside]{sfm}

\input{sf.def}

\begin{document}

\def\Teff{\ensuremath{T_{\mathrm{eff}}}}
\def\logg{\ensuremath{\log g}}
\def\vmic{$\upsilon_{\mathrm{mic}}$}
\def\vmac{$\upsilon_{\mathrm{macro}}$}
\def\vsini{\ensuremath{{\upsilon}\sin i}}
\def\kms{$\mathrm{km\,s}^{-1}$}
\def\ms{$\mathrm{m\,s}^{-1}$}
\def\exc{$\chi_{\mathrm{excit}}$}
\def\loggf{log$gf$}
\def\vr{${\upsilon}_{\mathrm{r}}$}
\def\logt{\ensuremath{\log t}}
\def\espa{ESPaDOnS}
\def\mus{MuSiCoS}
\def\nlte{non-LTE}
\def\llm{{\sc LLmodels}}
\def\atlas{{\sc ATLAS9}}
\def\tauros{\tau_{\rm ross}}
\def\logl{\ensuremath{\log L/L_{\odot}}}
\def\mbol{$M_{\mathrm{bol}}$}
\def\errvsini{$\sigma_{\ensuremath{{\upsilon}\sin i}}$}
\def\errvr{$\sigma_{{\upsilon}_{\mathrm{r}}}$}
\def\errTeff{$\sigma_{T_{\mathrm{eff}}}$}
\def\errlogg{$\sigma_{\ensuremath{\log g}}$}
\def\errvmic{$\sigma_{\upsilon_{\mathrm{mic}}}$}
\def\M{\ensuremath{M/M_{\odot}}}
\def\vald{{\sc VALD}}
\def\synth{{\sc SYNTH3}}

\pagebreak

\thispagestyle{titlehead}

\setcounter{section}{0}
\setcounter{figure}{0}
\setcounter{table}{0}

\markboth{Fossati, L.}{Chemical abundance studies of CP stars in open clusters}

\titl{Chemical abundance studies of CP stars in open clusters}{Fossati L.$^1$}
{$^1$Argelander-Institut f\"ur Astronomie der Universit\"at Bonn, Auf dem H\"ugel 71, 53121, Bonn, Germany, email: {\tt lfossati@astro.uni-bonn.de}}

\abstre{
In stellar astrophysics, the study of the atmospheres of early-type stars plays a very special role. The atmospheres of these stars display a variety of different phenomena, such as the presence of large magnetic fields, strong  surface convection, pulsation, diffusion of chemical elements. In particular, about 10\% of early F-, A- and late B-type stars present chemical peculiarities, which rise as a result of diffusion. A detailed study of the evolution of the chemical peculiarities as traces of diffusion processes requires the precise knowledge of stellar ages and initial chemical composition. Open clusters provide these information: 1) it is possible to assume that all cluster members have approximately the same original chemical composition and age; 2) the age of stars belonging to open clusters can be determined with much higher accuracy than for field stars. For this reason chemically peculiar stars member of open clusters have been targeted to study the evolution of the chemical peculiarities primarily to provide constraints to diffusion models. We review the abundance studies of chemically peculiar stars in open clusters performed until now, putting their results into the broader context of stellar evolution.
}

\baselineskip 12pt

\section{Introduction}
Stellar evolution is the study of how stars change with time and of the physical processes driving these changes. The study of stellar evolution is of crucial importance in almost every astrophysical area, from planets to galaxies. For example, our understanding of the past and future of the Earth's climate and habitability is tightly bound to our understanding of the Sun and of its evolution as a star. 

Stars spend most of their lifetime on the main sequence, burning hydrogen in the core, during which they go through only little structural changes. Main sequence stars are therefore well studied and believed to be well understood. Nevertheless, several physical processes such as diffusion, convection and rotation, driving stellar evolution along the main sequence phase, are far from being well understood and reproduced by the current state-of-the-art stellar evolution models (e.g., Langer~\cite{langer2012}). The study of chemically peculiar (CP) stars allows us to better understand some of the most basic physical processes driving stellar evolution, such as diffusion.

In absence of mixing (e.g., no rotation and convection) the diffusion processes (see e.g., Michaud~\cite{michaud1970} and references therein), which lead to chemical peculiarities in CP stars, can be generally simplified as follows. Each ion inside a star is simultaneously pushed towards the outer regions by the pressure of the photons coming from the star's core (i.e., radiation pressure), and towards the stellar interior by the star's gravity. Following their basic structure and characteristics, such as their mass and cross-section to radiation, different ions react differently. In a perfectly hydrostatic stellar envelope, the ions for which radiation pressure is stronger then gravity will be pushed towards the star's surface leading to an observable overabundance of the considered ions; on the other hand, the ions, for which gravity is stronger than radiation pressure, will sink inside the star, leading to an observable underabundance in the star's envelope (e.g., Richer et al.~\cite{richer2000}).

Since diffusion is a time-dependent process involving the evolution of chemical abundances, the knowledge of the stellar age and initial chemical composition is crucial. Open cluster stars are therefore the best suitable targets to study diffusion as a function of time and stellar mass, because of two very compelling reasons. First, it is possible to assume that all cluster members have approximately the same original chemical composition and age. Therefore, when the analysis of chemical abundances of stars belonging to the same cluster is performed, it is possible to assume that the studied objects are different only by their initial mass, rotational velocity and binarity. Second, the age of stars belonging to open clusters can be determined with much higher accuracy then for objects in the field. This is clearly shown in Fig.~1 of Bagnulo et al.~\cite{bagnulo2006}: assuming for a given field early-type star the effective temperature (\Teff) and luminosity (\logl) are know with typical uncertainties of 5\% and 0.1\,dex, respectively, one can derive the star's age with an average uncertainty of 25\%. In particular, for stars younger than half of their main sequence lifetime the age is completely unconstrained. This consideration does not include the additional uncertainty on the metallicity (initial chemical composition). On the other hand, for open cluster stars the age is that of the cluster and it can be determined with an uncertainty of 8\% or less, regardless of the cluster age. In addition, from the analysis of the chemically normal cluster stars one can derive the initial chemical composition.
\subsection{Determining cluster memberships}
In most cases, one of the major challenges of studying cluster stars is determining their membership. Cluster membership probabilities can be determined directly, using photometry and velocities (i.e., proper motion and radial velocity), or indirectly, with age sensitive features (e.g., emission lines and lithium abundance)\footnote{Note that parallaxes can also be used to determine cluster membership probabilities, but they are currently available only for a few open clusters (van Leeuwen~\cite{leeuwen2009}).}. We will now consider only the direct methods. 

Within the uncertainties, stars member of the same cluster lie on the isochrone corresponding to the cluster age and metallicity. Multicolor photometry allows one to determine on the color-magnitude diagram the distance between each observed star and the best fitting cluster isochrone, leading therefore to cluster membership probabilities. It is strongly recommended to apply this procedure also to the color-color diagram as some stars will appear as cluster members only in one diagram. More on the importance of using both color-magnitude {\it and} color-color diagrams can be found in Dias et al.~\cite{dias2012}.

\begin{figure}[!t]
\begin{center}
\hbox{
 \includegraphics[width=5.5cm]{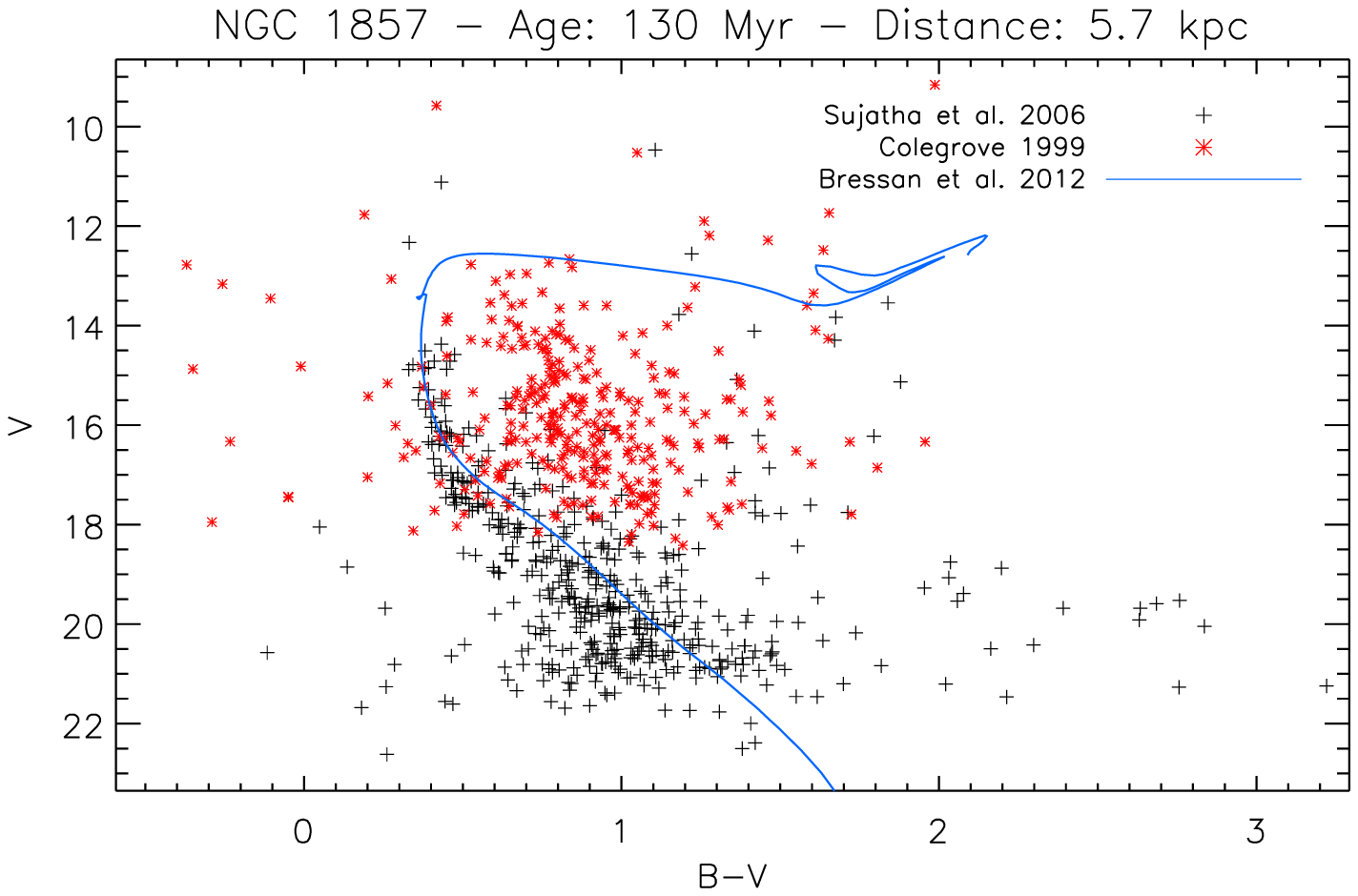}
 \includegraphics[width=5.5cm]{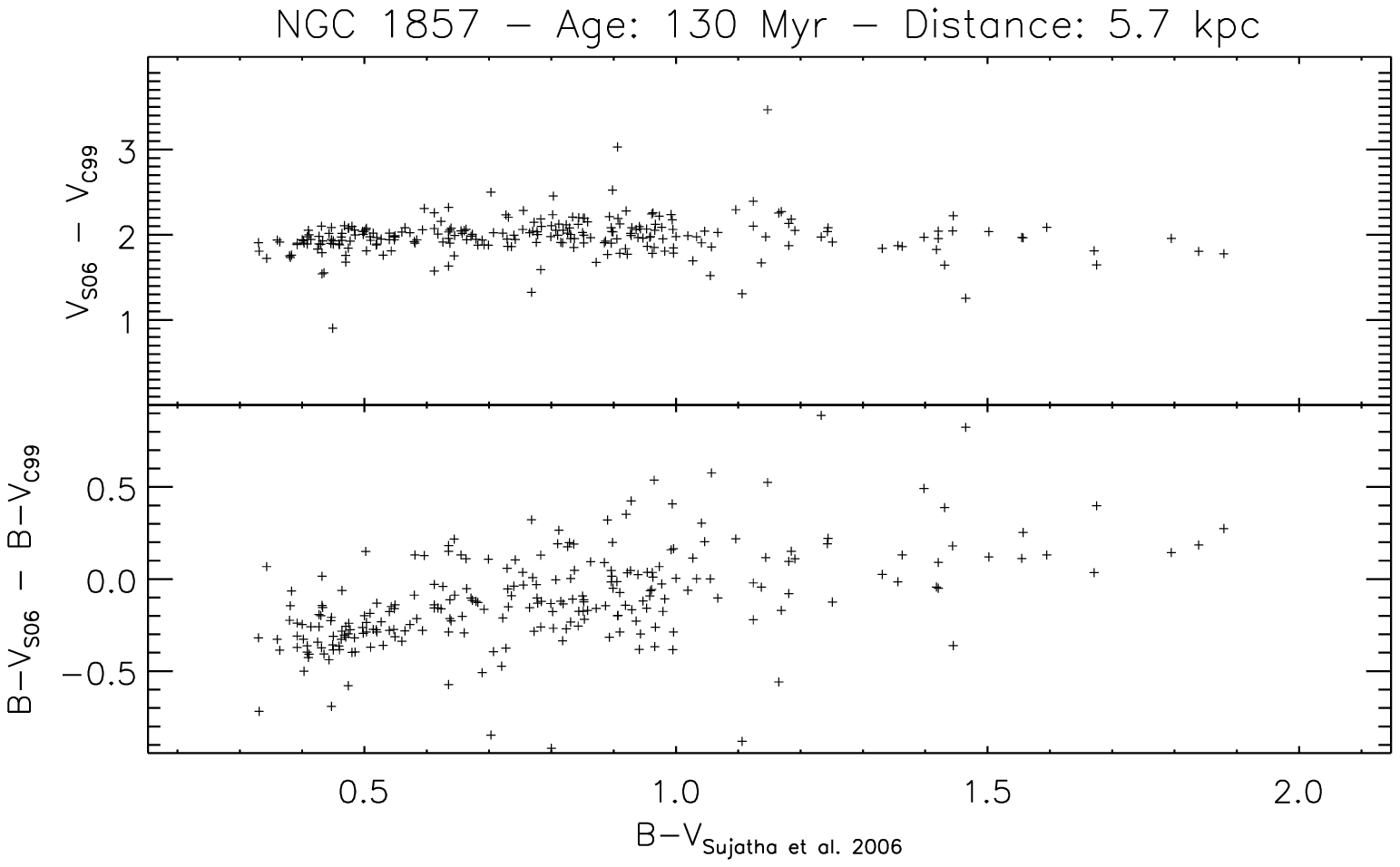}
}
\vspace{-5mm}
\caption[]{Left panel: comparison between the color magnitude diagram, resulting from CCD photometry with the Johnson $B$ and $V$ filters, of the open cluster NGC\,1857 as obtained by Sujatha et al.~\cite{ngc1} and Colegrove~\cite{ngc2}. Data taken from the WEBDA database. The continuous line shows the cluster best fitting isochrone, from Bressan et al.~\cite{bressan2012}. Top-right: comparison between the Johnson $V$ band magnitudes obtained by Sujatha et al.~\cite{ngc1} and Colegrove~\cite{ngc2} as a function of the color $B-V$ by Sujatha et al.~\cite{ngc1}. Bottom-right: as for the top-right panel, but for the color $B-V$.}
\label{membership1}
\end{center}
\end{figure}

Inhomogeneities of photometric catalogs for the same cluster are a further problem which can be encountered when determining cluster membership probabilities through photometry. As a matter of fact, looking at the open cluster photometric data available through the WEBDA database (Mermilliod \& Paunzen~\cite{webda}) one can notice large discrepancies between the magnitudes obtained by different authors on the same cluster, though using the same filters. Figure~\ref{membership1} shows, as an example, the difference between the magnitudes obtained by Sujatha et al.~\cite{ngc1} and Colegrove~\cite{ngc2}  of stars in the field of view of the young open cluster NGC\,1857. Both data-sets were obtained from CCD photometry in the Johnson $B$ and $V$ filters. As shown in Figure~\ref{membership1}, the difference is as large as 2\,mag and its origin is unknown. Other open clusters (e.g., Berkeley\,80, NGC\,1513, NGC\,2112, NGC\,6451, NGC\,2425, NGC\,3590, NGC\,7419; Paunzen, E., priv. comm.) show similar discrepancies in the available photometry. For this reason, we suggest to use always all available photometric references for a membership probability determination.

\begin{figure}[!t]
\begin{center}
\hbox{
 \includegraphics[width=8.5cm]{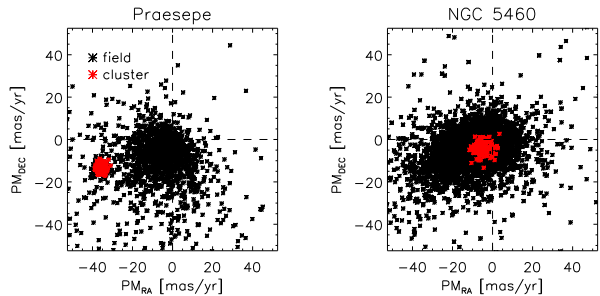}
}
\vspace{-5mm}
\caption[]{Left panel: proper motions (from Kharchenko et al.~\cite{kharchenko2004}) of stars lying at an angular separation smaller than 3.5 degrees from the center of the Praesepe open cluster. Right panel: proper motions of stars lying at an angular separation smaller than 10 degrees from the center of the open cluster NGC\,5460. Cluster members are marked in red.}
\label{pm}
\end{center}
\end{figure}

Proper motions, particularly for the most nearby clusters, are a very powerful tool to determine cluster membership probabilities. The left panel of Fig.~\ref{pm} shows the proper motions (from Kharchenko et al.~\cite{kharchenko2004}) of stars lying at an angular separation smaller than 3.5 degrees from the center of the Praesepe open cluster. It is evident that the cluster stars have a peculiar proper motion compared to that of the field stars, therefore allowing one to determine rather precise and unambiguous membership probabilities. On the other hand, the majority of the open clusters have a proper motion similar to that of the field stars (see for example the right panel of Fig.~\ref{pm}). 

Radial velocities (\vr) can also be used as cluster membership indicators, but precise average \vr\ values are known only for a few clusters. When using this indicator, one should always check how many stars were used to determine the average cluster \vr\ and how many times each star has been observed, in order to reject the binaries. This is particularly important for young cluster, were the brightest stars are very massive (earlier than B5) and therefore likely to be in a binary system (Sana et al.~\cite{sana}).

Given the various issues described above, we suggest to always use all available information and tools to determine cluster membership probabilities, in a similar fashion to what presented by Kharchenko et al.~\cite{kharchenko2005}, but including more photometric data sources and carefully checking the average cluster \vr\ against binarity.
\section{Elemental abundances of open cluster chemically peculiar stars}
The standard classification of CP stars is that given by Preston~\cite{preston} who subdivided CP stars into four groups: metallic line stars (CP1, otherwise called Am stars), magnetic Ap stars (CP2), HgMn stars (CP3), and He-weak stars (CP4). Magnetic fields have been detected and confirmed only for the CP2 and CP4 stars, while the non-magnetic CP stars (CP1 and CP3) usually belong to binary systems and rotate slower than stars with a normal (i.e., close to Solar) chemical composition (Abt~\cite{abt}). 

In the following we will review the results obtained from homogeneous chemical abundance studies of CP stars in open clusters, subdividing them on the basis of the chemical peculiarity: metallic line stars (Sect.~\ref{Am}), HgMn stars (Sect.~\ref{HgMn}), and magnetic CP stars (Sect.~\ref{magnetic}). Section~\ref{Am} includes also results obtained for open cluster chemically normal stars, in the context of diffusion.
\subsection{Chemically normal and metallic line stars}\label{Am}
Metallic line stars (CP1) have masses between about 1.2 and 3.0\,\M\  (early F- and A-type stars) and are characterised mostly by an underabundance of Sc (up to 2\,orders of magnitude), and overabundances of Fe-peak and rare-earth elements (up to 4\,orders of magnitude - see e.g., Fossati et al.~\cite{fossati2007}). In addition, they show large microturbulence velocity values, usually around 4\,\kms, compared to $\sim$2\,\kms\ typical of chemically normal stars (Landstreet et al.~\cite{landstreet2009}).

The first to systematically analyse and compare the abundance pattern of CP1 and chemically normal stars in open clusters were Hui-Bon-Hoa et al.~\cite{hui1997}, Hui-Bon-Hoa \& Alecian~\cite{hui1998}, and Hui-Bon-Hoa~\cite{hui1999}. Burkhart \& Coupry~\cite{BC2000} looked for systematic differences in the lithium abundance between open cluster CP1 and chemically normal stars, concluding that lithium is depleted by 3 times in CP1 stars.

Varenne \& Monier~\cite{VM} were instead the first to look for correlations between abundances and stellar parameters. For the chemically normal stars member of the Hyades open cluster, they found an anti-correlation between Fe abundance and projected rotational velocity (\vsini) for stars with \vsini$>$100\,\kms. A later inspection of the spectra revealed this finding was due to a wrong continuum normalisation (Monier R., priv. comm.). This was further confirmed by Gebran et al.~\cite{gebran2010}, who also concluded that CP1 chemical peculiarities result to be present up to \vsini$\sim$100\,\kms. The absence of correlations between element abundance and \vsini\ for both chemically normal and CP1 stars was also reported by Gebran et al.~\cite{gebran2008coma}.

Fossati et al.~\cite{fossati2008} analysed a sample of CP1 stars belonging to the Praesepe open cluster. In contrast to the previous results, they obtained that the abundances of the elements peculiar in CP1 stars display a significant correlation with \vsini, with peculiarities decreasing with increasing \vsini. This result is in agreement with diffusion model calculations by Talon et al.~\cite{talon2006}. The reason for the discrepancy between the analysis of the different clusters is unknown, though the most realistic explanation is differences in the adopted analysis method.

Direct comparisons between diffusion models and observed abundance patterns of open cluster CP1 stars have been presented by Gebran et al.~\cite{gebran2008coma},~\cite{gebran2010}, and Vick et al.~\cite{vick2010}. The major common conclusion is that diffusion models are able to reproduce either the light elements (i.e., atomic number $\lesssim$15) or the heavier elements (i.e., Fe-peak elements). In addition, there is always a significant discrepancy between the observed sodium abundance in CP1 stars compared to that predicted by diffusion models, with the observed abundance being systematically higher. Modelling suggests that sodium should not be affected by diffusion, whereas observationally Takeda et al.~\cite{takeda} concluded that sodium behaves like iron and therefore it should be treated as a ``peculiar'' element.

By looking at the whole set of stars (CP and normal) analysed by Monier~\cite{monier} in the Uma group and by Gebran \& Monier~\cite{gm} in the Pleiades, the authors found that the spread in the Fe abundance decreases with decreasing temperature. The authors suggested that rotational mixing in the radiative region might be the cause of this behavior. 

A further result of the analysis of open cluster stars is that diffusion appears to be efficient also in chemically normal stars of the upper main-sequence, regardless of their rotation rate. Fossati et al.~\cite{fossati2011} derived the abundance for a number of elements in F-, A- and B-type stars belonging to the NGC\,5460 open cluster. They found that the abundance of Mg and Fe correlates with \Teff, with an abundance increase for temperatures up to \Teff$\sim$10500\,K and a rather steep abundance decrease for higher temperatures. 
After excluding systematic effects, they concluded that diffusion is the most plausible cause of this behavior.
\subsection{Mercury-manganese stars}\label{HgMn}
Mercury-manganese stars (CP3) lie between about 2.5 and 5\,\M\  (late B-type stars) and are characterised mostly by a large overabundance of Mn and Hg (up to several orders of magnitude). In addition, CP3 stars present time-variable surface abundance spots (e.g., Kochukhov et al.~\cite{khochukhov2007}). 

There are no systematic abundance studies of cluster CP3 stars. The most relevant work is that presented by Wolf \& Lambert~\cite{WL} who discovered three CP3 stars in the Orion nebula, placing therefore an observational upper limit of 1.7\,Myr on the timescale needed to produce CP3 chemical peculiarities.
\subsection{Magnetic CP stars}\label{magnetic}
Magnetic CP stars can be found all along the upper main sequence and are characterised by the presence of sometimes complex and strong surface magnetic fields and various flavors of abundance anomalies (see e.g., Ryabchikova~\cite{tanya}). 

Only very recently systematic abundance studies of cluster magnetic stars have been published, but mostly devoted to the analysis of single stars (e.g., Bailey et al.~\cite{B1}, \cite{B2}, Bailey \& Landstreet~\cite{B3}). Nevertheless, at this conference, John Landstreet showed the first results obtained from the analysis of several magnetic CP stars in open clusters of different age. The results indicate that the chemical peculiarities decrease with increasing age (see Landstreet~\cite{john} and Bailey et al.~in prep. for more details).
\section{Conclusions}
The ESA GAIA satellite will soon provide us extremely precise distances for most of the stars we observe in the sky. For this reason it is legitimate to ask ourselves the following questions. Do we still need to study open cluster stars after GAIA? Bagnulo et al.~\cite{bagnulo2006} concluded that one can derive the star's main sequence lifetime with an average uncertainty of 25\%: is this still true after GAIA?

To answer these questions we repeat the experiment performed by Bagnulo et al.~\cite{bagnulo2006} in their Fig.~1 and assuming the expected very best case scenario \Teff\ and \logl\ uncertainties after GAIA. Although probably unrealistic, particularly for CP stars, some (e.g., Liu et al.~\cite{liu}) suggest that GAIA spectrophotometry would allow one to determine \Teff\ with a precision of 1\%. Regarding the luminosity, we can assume that after GAIA the uncertainty on \logl\ will be due only to the uncertainty on the bolometric correction and therefore we consider an uncertainty on \logl\ of 0.05\,dex.

\begin{figure}[!t]
\begin{center}
\hbox{
 \includegraphics[width=8.5cm]{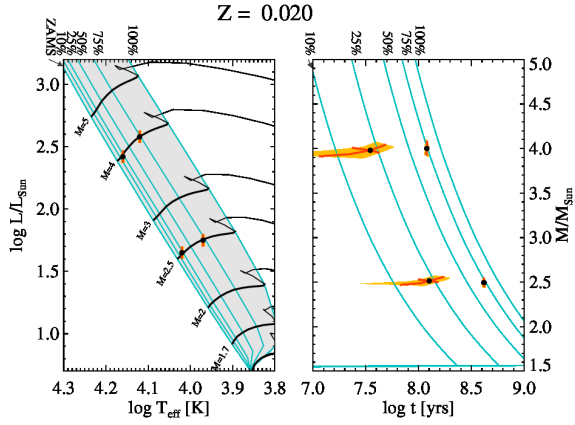}
}
\vspace{-5mm}
\caption[]{Position of a star in the HR diagram (left panel), and the star's position transformed into a diagram of age as a function of stellar mass (right panel), assuming one knows \Teff$\pm$1\% and luminosity $\pm$0.05\,dex. The transformation uses standard evolution tracks for $Z$ = 0.02
(Schaller et al.~\cite{schaller}); several fractional ages (fraction of main sequence life completed) are labelled. Figure made by Vincenzo Andretta.}
\label{hr}
\end{center}
\end{figure}

Figure~\ref{hr}, to be compared to Fig.~1 of Bagnulo et al.~\cite{bagnulo2006}, was produced assuming these uncertainties. Even with the very best case scenario, it will not be possible to determine the age of young field stars with the same precision of that given by open clusters. More importantly only open clusters will provide the crucial information on the initial chemical composition, impossible to derive for field CP stars.

%
\bigskip
{\it Acknowledgements.} The author thanks Vincenzo Andretta for making Figure~\ref{hr}. This work made use of the WEBDA database.

\end{document}